\documentclass[aps,twocolumn,showpacs,floatfix]{revtex4-1}
\usepackage{dcolumn}
\usepackage{bm}
\usepackage{amsmath}
\usepackage{soul}
\usepackage{amssymb}
\usepackage{epsfig}
\usepackage{tikz}
\usepackage{xcolor}
\usetikzlibrary{shapes,chains,positioning,snakes}
\usepgflibrary{snakes}

\begin{document}
\title{Finite-time erasing of information stored in fermionic bits}
\author{Giovanni Diana $^1$}
\author{G. Baris Bagci $^2$}
\author{Massimiliano Esposito $^1$}
\affiliation{
$^1$ Complex Systems and Statistical Mechanics, University of Luxembourg, L-1511 Luxembourg, Luxembourg\\
$^2$ Department of Physics, Faculty of Science, Ege University, 35100 Izmir, Turkey
}


\begin{abstract}
We address the issue of minimizing the heat generated when erasing the information stored in an array of quantum dots in finite time. We identify the fundamental limitations and trade-offs involved in this process and analyze how a feedback operation can help improve it.      
\end{abstract}


\maketitle

\section{Introduction}

Establishing the thermodynamic cost of various operations processing information is a fundamental issue technologically as well as conceptually. Its origin can probably be traced back to Maxwell discussing his well known demon \cite{Leff}. An important result in this respect was achieved by Landauer who established a lower bound for the heat generated when erasing one bit of information \cite{Landauer61}. In doing so, he established an explicit connection between a thermodynamic quantity, heat, and a quantity measuring information, the Shannon entropy \cite{Shannon48}. The validity of this so-called Landauer principle has been verified in the context of classical, quantum and stochastic dynamics \cite{Piechocinska00, Shizume1995, EspoVdB_EPL_11}. Its relation to entropy production and microscopic reversibility was discussed in Refs.\cite{Gaspard04b, VandenBroeck07, GaspardAndrieuxEPL08}. The implications of Landauer's principle for computation were recognized early on \cite{Landauer70, Bennett73, 82BennettThermoComp} and opened the way to the field of reversible computing \cite{DeVos}.

With the advent of stochastic thermodynamics \cite{Seifert12Rev,Esposito12}, Landauer's principle has become an immediate consequence of the second law. In this formalism, a system consists of states $i$ with energies $\epsilon_i$ and probabilities $p_i$. The Shannon entropy of the system is given by $S=-k_B \sum_i p_i \ln p_i$. When the system is in contact with a single reservoir at temperature $T$, transitions between the system states occur and the system probabilities evolve according to the Markovian master equation $\dot{p}_i=\sum_j w_{ij} p_j$ with transition rates $w_{ij}$ which satisfy the local detailed balance condition $w_{ij}/w_{ji}=\exp{\{(\epsilon_j-\epsilon_i)/(k_B T)\}}$. The second law in stochastic thermodynamics reads $\Delta S=Q/T+\Delta_{\rm \bold i} S$, where $\Delta S$ is the change in Shannon entropy, $Q$ is the integrated heat flow $\dot{Q}=\sum_i \epsilon_i \dot{p}_i$ entering the system, and $\Delta_{\rm \bold i} S$ is the nonnegative entropy production which only vanishes for quasi-static transformations where detailed balance is satisfied. If the system is a bit (i.e. a two level system with two states $0$ and $1$) initially containing the maximal information $S=k_B \ln 2$ corresponding to the uniform probability $p_0=p_1=1/2$, Landauer's principle states that erasing that information, i.e. bringing the initial system entropy to $S=0$, will produce an amount of heat of at least $k_B T \ln 2$. This immediately follows from the second law since $\Delta S=k_B \ln 2$ and therefore the generated heat reads $-Q=k_B T \ln 2+\Delta_{\rm \bold i} S \geq k_B T \ln 2$. Landauer's lower bound is only reached for quasi-static transformations where $\Delta_{\rm \bold i} S=0$ and thus requires an infinite amount of time.

Since stochastic thermodynamics naturally combines dynamics with thermodynamics, it opens the way to the study of information erasure in finite-time. Interesting results in this direction have been obtained for systems described by Fokker-Plank equations. Transformations of duration $t$ between two sets of probabilities which minimize the heat generated lead to an entropy production scaling as $1/t$ \cite{Seifert07b, AurellGawedzki12}. This result also holds for systems described by master equations but is limited to a regime of low dissipation \cite{EspKawLindVdBEPL10}. These studies have important implications for the study of efficiencies in finite-time thermodynamics \cite{SeifertSchmiedlPRL07, Seifert07b, SeifertMarinJCP08, 10EspoKawLindVDB_PRL, EspoKawLindVdB_PRE_10}. Furthermore, many recent works have analyzed the implications that feedback control may have on the thermodynamic description of a system \cite{SagawaUedaPRL08, SagawaUedaPRL09, SagawaUedaPRL10, Horowitz10, SeifertEPL11, SeifertAbreuPRL12, JarzynskyPNAS12, EspositoSchaller12, EspositoStrassSchall12, ParrondoHoroSag12}. In this paper we are going to build on these studies to analyze the process of information erasure in finite time first without and then in the presence of feedback control.

In the first part of this paper, i.e. section \ref{section:one}, we introduce our model and study in detail the erasure of information in finite time by analyzing the trade-offs between generated heat, erasure time, and accuracy of the erasure. We also introduce the notion of erasing efficiency and erasing power. In the second part of the paper, i.e. section \ref{section:two}, we study how to improve the erasure process by introducing a feedback process. Conclusions are drawn in section \ref{section:conclu}.    

\section{Erasing information in finite time} \label{section:one}

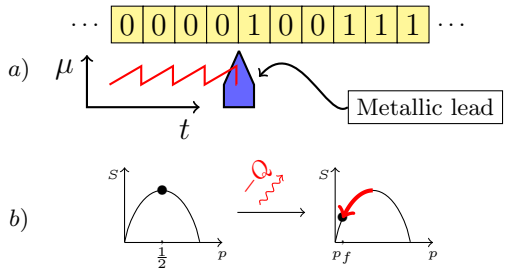
\begin{figure}[h!]
\begin{tikzpicture}[
      start chain=1 going right,start chain=2 going below,node distance=-0.15mm, start chain=3 going right,segment length=4pt,segment amplitude=1pt
    ]
    \node at (-2.4,-1) {$a)$};
    \node [on chain=2] {};
    \node [on chain=1,name=l] at (-1.5,-.4) {\ldots};  
    \foreach \x in {0,0,0,0,1,0,0,1,1,1} {
        \x, \node [draw,on chain=1,fill=yellow!50] {\large \x};
    } 
    \node [name=r,on chain=1] {\ldots}; 
    \node [on chain=3] at (-1.5,-1.) {}; 
    \node [on chain=3] {};   
    \node[on chain=2] (start2) at (0.7,-.75) {};
     \draw[thick,fill=blue!60,xshift=.6] (0.5,-.75) --  (0.7,-1.2) -- (0.7,-1.5) -- (0.3,-1.5) -- (0.3,-1.2) -- (0.5,-.75); 
    \draw[snake=saw,color=red,thick,segment length=12pt,segment amplitude=7pt] (-1.2,-1.2)-- (0.49,-1.2);
    \node at (1.5,-.85) {};
    \node [on chain=2] {};
    \node [on chain=2] {};
    \node [on chain=2] {};
    \node (mu) at (-1.8,-1) {\large $\mu$};
    \node (t) at (-.2,-1.8) {\large $t$};
    \draw[<->,shorten <=2ex,thick]  (l) -- (-1.5,-1.5) -- (0,-1.5);
    
    \node [draw] (ml) at (3,-1.5) {Metallic lead};
    \path [->,draw,thick,shorten >=1ex] (ml.west) ..controls (1.7,-1.7) and (1.3,-.5) .. (0.7,-1.2);
    \node at (-2.4,-3.0) {$b)$};
    \begin{scope}[shift={(-1,-3.3)}]
       \draw[->] (0,0) -- (1.3,0)  node[below] {\tiny $p$};
       \node at (0.5,0) {\tiny ,};
       \node at (0.5,-.2) {\tiny$\frac{1}{2}$};
       \draw[->] (0,0) -- (0,1); 
       \node at (-.15,.9) {\tiny $S$};
       \draw[domain=0.001:.999] plot (\x,{-\x*ln(\x)-(1-\x)*ln(1-\x)});
       \node at (.5,.693) {$\bullet$};
    \end{scope}
    \draw[->] (0.5,-2.9) -- (1.4,-2.9);
    \draw[snake=snake,->,color=red] (0.8,-2.8) -- (1.1,-2.4) node[midway,sloped,above] {\small$-Q$};
    \begin{scope}[shift={(1.8,-3.3)}]
       \draw[->] (0,0) -- (1.3,0)  node[below] {\tiny $p$};
       \draw[->] (0,0) -- (0,1); 
       \node at (-.15,.9) {\tiny $S$};
       \node at (.1,0) {\tiny ,};
       \node at (.1,-.2) {\tiny$p_f$};
       \node at (.1,0.325) {$\bullet$};
       \draw[domain=0.001:.999] plot (\x,{-\x*ln(\x)-(1-\x)*ln(1-\x)});
       \draw[domain=0.1:.5,<-,color=red,line width=.4ex] plot (\x,{-\x*ln(\x)-(1-\x)*ln(1-\x)});
    \end{scope}

\end{tikzpicture}
\caption{(Color online) 
a) The information stored in the array of single level quantum dots is erased by putting the quantum dots one after another in contact with the metallic lead and applying, during the contact time $t$, a time dependent protocol on the lead chemical potential $\mu(t)$.  
b) The decrease of the initial Shannon entropy $S_i=k_B \log2$ is accompanied by an heat release $-Q$ in the environment.\label{fig:dotdevice}} 
\end{figure}

We consider classical information stored in an array of single level quantum dots. Each dot constitutes a classical bit since it can either be empty or filled with an electron ($0$ or $1$) with probability $1-p$ and $p$ respectively. The Shannon entropy per bit is 
\begin{equation}
S=-k_B p \ln p - k_B (1-p) \ln (1-p). 
\end{equation}
It takes its maximal value $S=k_B \ln 2$ when $p=1/2$ and its minimal value $S=0$ when $p=0$ or $p=1$. The energy of an electron in the dot is denoted by $E$. The stored information is metastable in the sense that the energy gap $E-\mu_{\rm env}$ to bring an electron in or out of the dot from or in its surrounding environment is much larger than the energy fluctuations $k_B T$ of the environment. However, it can be modified when the dot enters in contact with a metallic lead moving at constant speed along an array of quantum dots as depicted in Fig.~\ref{fig:dotdevice}. 
The speed of the lead controls the contact time $t$ between the dot and the lead. The lead is at the surrounding temperature $T$ but its chemical potential $\mu(t)$ is externally controlled. Its time dependence during the contact time $t$ will be denoted as the protocol and is the same for each dots. The dynamics of the dot during the contact time is described by the master equation
\begin{equation} \label{eq:mastereq}
\dot p=- C (1-f(E)) p + C f(E) (1-p),
\end{equation}
where $C f(E)$ (resp. $C (1-f(E))$) is the rate at which the lead can donate (resp. receive) an electron to (resp. from) the dot, and $f(E)=(\exp{\{(E-\mu)/(k_BT)\}}+1)^{-1}$ is the Fermi distribution of the lead. Introducing the variable $\epsilon=E-\mu$, setting $k_B=1$, and measuring time in units of $C^{-1}$, we can rewrite (\ref{eq:mastereq}) as
\begin{equation} \label{eq:model}
\dot p=-p+\frac{1}{\exp{\{\epsilon(t)/T\}}+1}.
\end{equation}
When solving this equation over the contact time $t$ with a time dependent protocol $\epsilon(t)$, the probability $p$ evolves from an initial value $p_i=p(0)$ to a final value $p_f=p(t)$. The resulting change in Shannon entropy per bit is given by the second law of stochastic thermodynamics \cite{EspoVdB10_Da}
\begin{equation} \label{eq:2law}
\Delta S = S_f-S_i =\frac{Q}{T} + \Delta_{\rm \bold i} S,
\end{equation}
where $S_f$ and $S_i$ are the Shannon entropies corresponding to $p_f$ and $p_i$ respectively. The heat entering the dot is given by
\begin{equation} \label{eq:Qintegral}
Q = \int_0^t d\tau \; \dot{p}(\tau) \epsilon(\tau)
\end{equation}
and the resulting nonzero entropy production reads
\begin{equation} \label{EP}
\Delta_{\rm \bold i} S = \int_0^t d\tau \; \dot{p}(\tau) \big( \ln \frac{1-p(\tau)}{p(\tau)} - \frac{\epsilon(\tau)}{T} \big) \geq 0.
\end{equation}

An erasure process is characterized by a negative entropy change $\Delta S<0$. Due to Eq.~(\ref{eq:2law}), this process releases heat into the environment (i.e. the heat absorbed is negative $Q<0$). We define the efficiency of this process as the amount of entropy change resulting from this heat release 
\begin{equation}\label{eq:eta}
0 \leq \eta = \frac{-\Delta S}{-Q/T} = 1 - \frac{\Delta_{\rm \bold i} S}{-Q/T} \leq 1.
\end{equation}
The upper bound of the efficiency $\eta=1$ corresponds to Landauer's lower bound and is reached when the protocol evolves quasistatically between $\epsilon(0)=T \ln (1/p_i-1)$ and $\epsilon(t)=T \ln (1/p_f-1)$ so that the entropy production vanish $\Delta_{\rm \bold i} S=0$. In this case the erasure process becomes infinitely slow. The lower bound $\eta=0$ will be reached when the heat generated for a given erasure diverges. 

Since the second term in Eq.~(\ref{eq:model}) is bounded between 0 and 1, the fastest way to decrease (resp. increase) the probability $p$ is given by $\dot p(t) = -p(t)$ (resp. $\dot p(t) = 1-p(t)$). This corresponds to a protocol $\epsilon \to \infty$ (resp. $\epsilon \to - \infty$) which leads to a divergent heat production and thus a vanishing efficiency $\eta=0$. This argument also implies that the time required to bring the probability from $p_i=p(0)$ to $p_f=p(t)$ is always larger or equal to a minimal time $t_\mathrm{min}$: 
\begin{eqnarray}
&&{\rm if} \; p_i>p_f  : \ \  t \ge t_\mathrm{min} = \ln \frac{p_i}{p_f}, \nonumber \\
&&{\rm if} \;  p_f>p_i : \ \  t \ge t_\mathrm{min} = \ln \frac{1-p_i}{1-p_f}. \label{eq:time}
\end{eqnarray}
The case $t=t_\mathrm{min}$ corresponds to the protocol $\epsilon \to \infty$ which leads to $\eta=0$. If $t<t_\mathrm{min}$, no protocol is able to reach $p_f$ from $p_i$. 
This brings us to the important result that perfect erasure ($S_f=0$ due to $p_f=0$ or $p_f=1$) leads to a divergent $t_\mathrm{min}$. In other words it is impossible to completely erase an initial finite Shannon entropy per bit in a finite amount of time. Only quasistatic process can do so, since in this case the contact time and the minimal time can diverge preserving $t \ge t_\mathrm{min}$. 

We can of course reformulate (\ref{eq:time}) by saying that for a finite contact time $t$, the final probability $p_f$ which defines the erasure error $S_f$ is always bounded by a critical probability $p_c$:
\begin{eqnarray}
&&{\rm if} \; p_i>p_f  : \ \  p_f \geq p_c = p_i \exp{\{-t\}}, \nonumber \\
&&{\rm if} \;  p_f>p_i : \ \ p_f \leq p_c = 1-(1-p_i) \exp{\{-t\}}.  \label{CritProb}
\end{eqnarray}
As a result, minimizing the erasing error for a given contact time corresponds to the case $p_f=p_c$ which leads to a divergent heat and thus to a vanishing efficiency $\eta=0$.

\begin{figure}[t]
\centering
\includegraphics[scale=0.27]{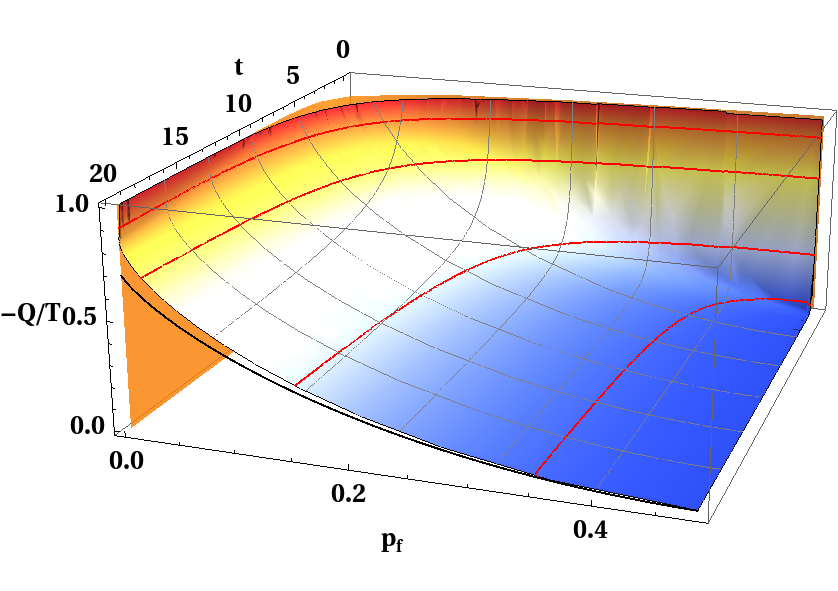}
\vspace{-0.3cm}
\caption{(Color online) Minimal heat generated $-Q/T$ when erasing in a finite time $t$ an initial information $S_i=T \ln 2$ ($p_i=1/2$) with a remaining error $S_f$ ($p_f$). The horizontal (red) curves on the surface correspond to $Q/T=0.05,\;0.3,\;\log2,\;0.8$. The vertical plane (orange) corresponds to $p_f=p_c$ where the heat diverges. The (black) curve in the $t=20$ plane corresponds to Landauer's lower bound where $Q/T=\Delta S$.}
\label{fig:3Dplot_1}
\end{figure}

\begin{figure}[b]
\centering
\includegraphics[scale=0.25]{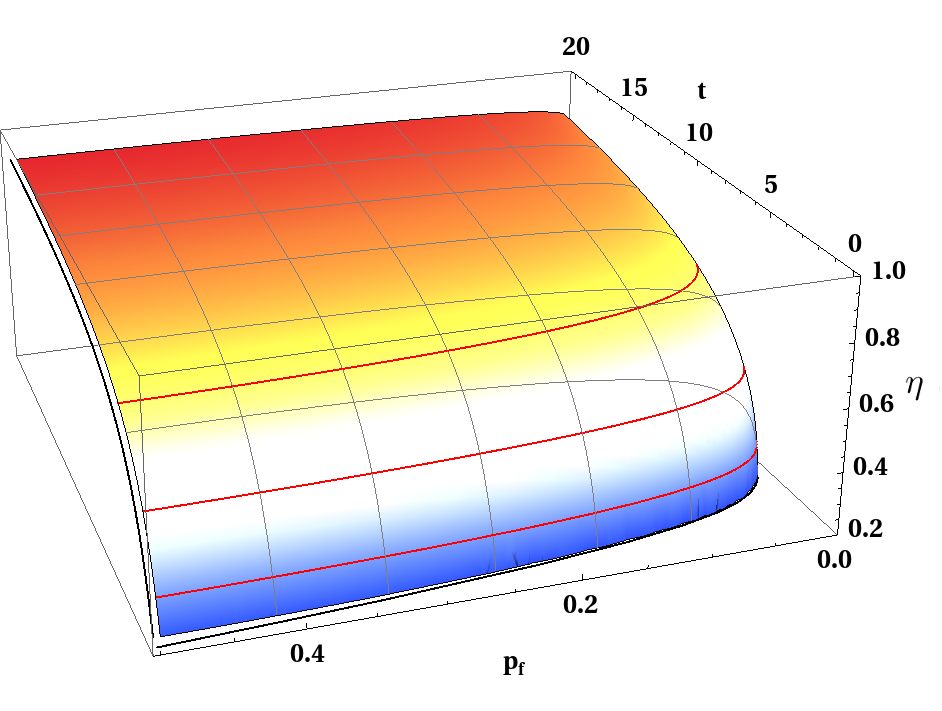}
\vspace{-0.3cm}
\caption{(Color online) Erasing efficiency $\eta$ corresponding to the erasure process in Fig.~\ref{fig:3Dplot_1}. The horizontal (red) curves on the surface correspond to $\eta=0.3,\;0.5,\;0.7$.}
\label{fig:3Dplot_3} 
\end{figure}

We now consider finite contact times $t$ with fixed $p_i$ and $p_f$ such that $p_f<p_c$ and try to find the protocol which maximizes the efficiency. Since $\Delta S$ is fixed, the highest efficiency will be obtained when the heat released in the environment is minimal or equivalently when the entropy production is minimal. The procedure to find the optimal protocol minimizing the heat released is detailed in Ref. \cite{EspKawLindVdBEPL10}. The amount of heat generated with the optimal protocol bringing the initial probability $p_i=1/2$ to the final value $p_f$ (corresponding to an erasing error $S_f$) in a time $t$ is displayed in Fig.~\ref{fig:3Dplot_1}. The corresponding erasing efficiency is depicted Fig.~\ref{fig:3Dplot_3}.
As the contact time increases and the protocol approaches the quasistatic solution, the heat approaches the Landauer limit $Q/T=\Delta S$ represented by the full (black) line in the $t=20$ plane of Fig.~\ref{fig:3Dplot_1} and the erasure efficiency increases and approaches one in Fig.~\ref{fig:3Dplot_3}. 
Also, as the final probability $p_f$ reaches its critical value $p_c$, the heat starts diverging and the efficiency drops to zero. This region where $p_f=p_c$ is represented in Fig.~\ref{fig:3Dplot_1} by a vertical (orange) plane and in Fig.~\ref{fig:3Dplot_3} by the full (black) line in the $\eta=0.2$ plane.
In the limit $p_f \to 1/2$, not surprisingly the heat vanishes, but the efficiency converges to the full (black) line in the $p_f=0.5$ plane of Fig.~\ref{fig:3Dplot_3} which can be calculated analytically using results of Ref.~\cite{EspKawLindVdBEPL10} and gives $\eta = (1+2/t)^{-1}$.
In Fig.~\ref{fig:3Dplot_1}, the behavior of $p_f$ as a function of $t$ for a fixed value of heat $Q$ is shown by horizontal (red) curves. The long time behavior of these curves depends on the value of the heat. For $-Q \leq T \ln 2$, it decreases to an asymptotic value of $p_f$ corresponding to the Landauer limit $\Delta S=S_f-\ln 2=Q/T$. However, for $-Q > T \ln 2$, it will eventually reach $p_f=0$.

\begin{figure}[h]
\centering
\includegraphics[scale=0.26]{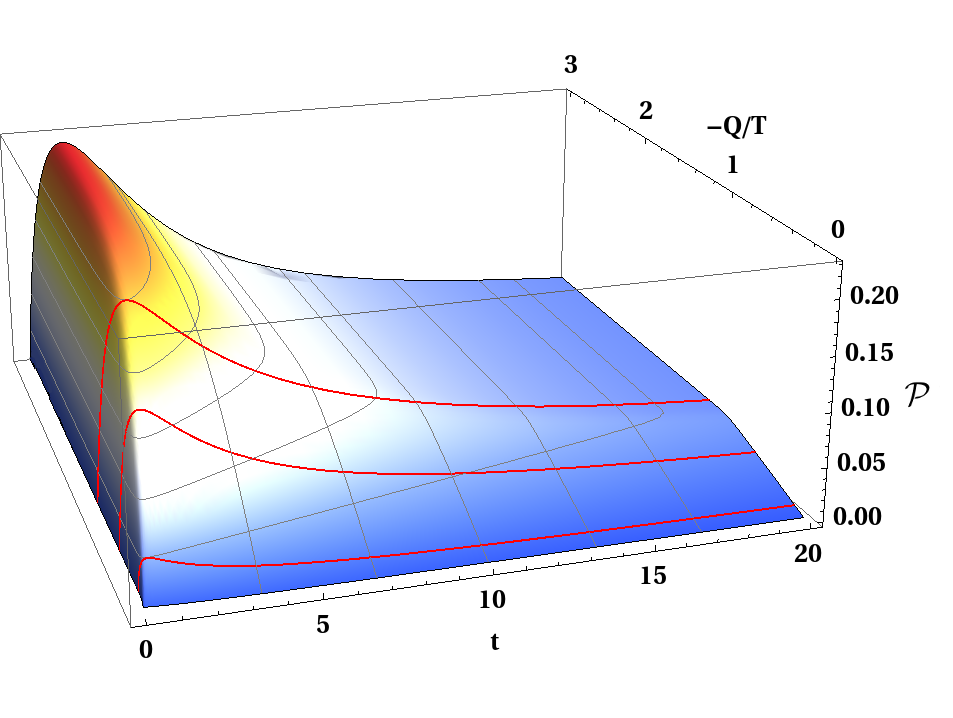}
\vspace{-0.7cm}
\caption{(Color online) Erasure power $\mathcal{P}$ as a function of heat generated $-Q/T$ and the duration of the erasure $t$. The (red) curves on the surface correspond to $Q/T=0.2,\;0.7,\;1.2$.}
\label{fig:3Dplot_2}
\end{figure}

In order to quantify the average amount of information erased per unit time during a contact time $t$ with a given amount of generated heat $-Q$, we define the erasure power 
\begin{equation}
\mathcal{P}(Q,t)=\frac{-\Delta S}{t},
\end{equation}
which is a function of $Q$ and $t$. The Landauer limit which leads to an optimal erasure efficiency ($\eta=1$) corresponds to zero erasure power ($\mathcal{P}=0$). Nonzero erasure power only occur at finite contact time as shown in Fig.~\ref{fig:3Dplot_2} where $\mathcal{P}$ is plotted as a function of the heat generated (with the optimal protocol minimizing heat) and time. 
We note that for a constant generated heat $-Q/T$, the erasure power $\mathcal{P}$ reaches a maximum value for relatively short times and then drops to zero in the long time limit where the optimal protocol becomes quasistatic. 

\begin{figure}[h!]
\centering
\includegraphics[scale=0.34]{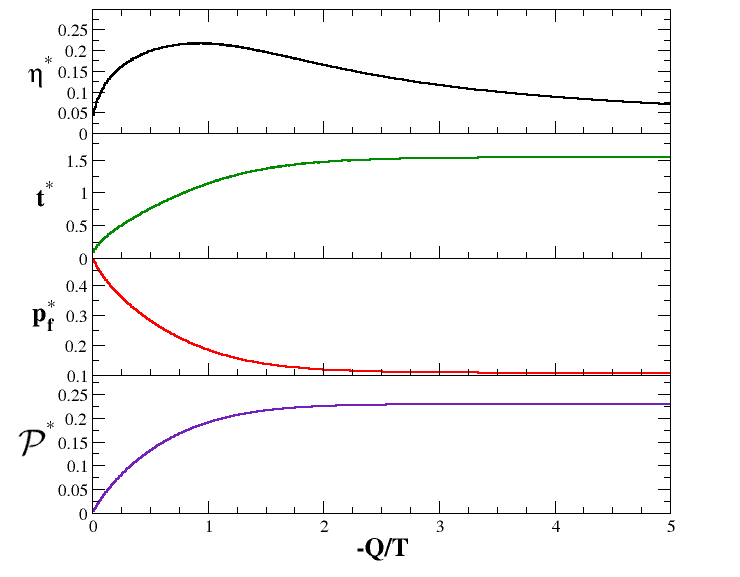}
\caption{(Color online) From upper to lower frame: Erasing efficiency at maximum erasing power (black), the corresponding contact time (green), final probability (red), and power (blue).}
\label{fig:effplot}
\end{figure}
The erasure efficiency at maximum erasure power $\eta^*$, as well as the corresponding contact time $t^*$, the corresponding final value of the probability $p_f^*$, and the corresponding value of the power $\mathcal{P}^*$ are displayed in Fig.~\ref{fig:effplot}. In the limit of small generated heat ($-Q/T \to 0$) the entropy change has to vanish since the second law imposes $-\Delta S \leq -Q/T$. It does so as $(1/2-p_f)^2$ since in that limit the final probability $p_f$ approaches $1/2$ which is a maximum of $\Delta S$. In the same limit the contact time can be shown to behave as $1/2-p_f$ so that the erasure power and the corresponding efficiency both vanish. In the opposite limit of large generated heat ($-Q/T \to \infty$), using (\ref{eq:time}), the erasure power behaves as $\mathcal{P}=-\Delta S/\ln(p_i/p_f)$ and its maximum value occurs at $p_f \approx 0.10892$. As a consequence the efficiency decreases in that limit as $\eta \propto Q^{-1}$. 

\section{Erasure with feedback} \label{section:two} 

\begin{figure}[h!]
\centering
\begin{tikzpicture}[
      start chain=1 going right,start chain=2 going below,node distance=-0.15mm, start chain=3 going right,segment length=4pt,segment amplitude=1pt
    ]
    \node at (-3.4,0) {$a)$};
    \node [on chain=2] {};
    \node [on chain=1,name=l] at (-1.5,-.4) {\ldots};  
    \foreach \x in {0,0,0,0,1,0,0,1,1,1} {
        \x, \node [draw,on chain=1,fill=yellow!50] {\large \x};
    } 
    \node [name=r,on chain=1] {\ldots}; 
    \node [on chain=3] at (-1.5,-1.) {}; 
    \node [on chain=3] {};   
    \node[on chain=2] (start2) at (0.7,-.75) {};
     \draw[thick,fill=blue!60,xshift=.6] (0.5,-.75) --  (0.7,-1.2) -- (0.7,-1.5) -- (0.3,-1.5) -- (0.3,-1.2) -- (0.5,-.75); 
    \draw[color=red,thick] (-1.2,-1.2) -- +(.45,.2)-- +(.45,0)--+(.87,.4)--+(.87,0)--+(1.27,.2)--+(1.27,0)--+(1.73,.4);
    \node at (1.5,-.85) {};
    \node [on chain=2] {};
    \node [on chain=2] {};
    \node [on chain=2] {};
    \node (mu) at (-1.8,-1) {\large $\mu$};
    \node (t) at (-.2,-1.8) {\large $t$};
    \draw[<->,shorten <=2ex,thick]  (l) -- (-1.5,-1.5) -- (0,-1.5);
    
    \node [draw] (ml) at (3,-1.5) {Metallic lead};
    \path [->,draw,thick,shorten >=1ex] (ml.west) ..controls (1.7,-1.7) and (1.3,-.5) .. (0.7,-1.2);
   \node (mic) at (.7,0.2) {\includegraphics[scale=.1]{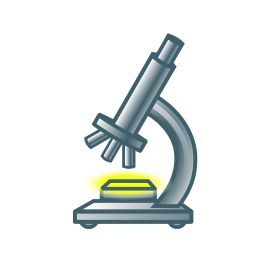}};
   \node[draw] (meas) at (1.7,.4) {Measure};
    \node at (-3.4,-3.0) {$b)$};
    \begin{scope}[shift={(-2.9,-4)}]
       \draw[->] (0,0) -- (1.3,0)  node[below] {\tiny$p$};
       \node at (0.5,0) {\tiny ,};
       \node at (0.5,-.3) {$\frac{1}{2}$};
       \draw[->] (0,0) -- (0,1);
       \node at (-.15,.9) {\tiny $S$};
       \draw[domain=0.001:.999] plot (\x,{-\x*ln(\x)-(1-\x)*ln(1-\x)});
       \node at (.5,.693) {$\bullet$};
    \end{scope}
    \draw[->] (-1.3,-3.3)-- (-0.5,-2.8) node [midway,sloped,above] {\tiny{$\begin{displaystyle}\bar\sigma=0\end{displaystyle}$}};
    \begin{scope}[shift={(-0.2,-3.3)}]
       \draw[->] (0,0) -- (1.3,0)  node[below] {\tiny $P(1|0)$};
       \node at (0.1,0) {\tiny ,};
       \draw[->] (0,0) -- (0,1);
       \node at (-.4,.9) {\tiny $S(\sigma|0)$};
       \node at (.1,0.325) {$\bullet$};
       \draw[domain=0.001:.999] plot (\x,{-\x*ln(\x)-(1-\x)*ln(1-\x)});
    \end{scope}
    \draw[->] (-1.3,-3.9)--(-0.5,-4.5) node[midway,sloped,below] {\tiny{$\begin{displaystyle}\bar\sigma=1\end{displaystyle}$}};;
     \begin{scope}[shift={(-0.2,-4.8)}]
       \draw[->] (0,0) -- (1.3,0)  node[below] {\tiny $P(1|1)$};
       \draw[->] (0,0) -- (0,1);
       \node at (-.4,.9) {\tiny $S(\sigma|1)$};
       \node at (0.9,0) {\tiny ,};
       \node at (.9,0.325) {$\bullet$};
       \draw[domain=0.001:.999] plot (\x,{-\x*ln(\x)-(1-\x)*ln(1-\x)});
    \end{scope}
     \begin{scope}[shift={(3,-3.3)}]
       \draw[->] (0,0) -- (1.3,0)  node[below] {\tiny $P(1|0)$};
       \draw[->] (0,0) -- (0,1) ;
       \node at (.05,0) {\tiny ,};
       \node at (.03,0.134) {$\bullet$};
       \draw[domain=0.001:.999] plot (\x,{-\x*ln(\x)-(1-\x)*ln(1-\x)});
       \draw[domain=0.03:.15,<-,color=red,line width=.4ex] plot (\x,{-\x*ln(\x)-(1-\x)*ln(1-\x)});
    \end{scope}
     \begin{scope}[shift={(3,-4.8)}]
       \draw[->] (0,0) -- (1.3,0)  node[below] {\tiny $P(1|1)$};
       \draw[->] (0,0) -- (0,1) ;
       \node at (.05,0) {\tiny ,};
       \node at (.03,0.134) {$\bullet$};
       \draw[domain=0.001:.999] plot (\x,{-\x*ln(\x)-(1-\x)*ln(1-\x)});
       \draw[domain=0.03:.9,<-,color=red,line width=.4ex] plot (\x,{-\x*ln(\x)-(1-\x)*ln(1-\x)});
    \end{scope}
    \draw[->] (1.6,-3.0)--(2.6,-3.0) ;
    \draw[snake=snake,->,color=red] (2.3,-2.9)--+(.2,.4);
    \node[color=red] at (1.9,-2.7){\tiny $-Q(0)$};
    \draw[->] (1.6,-4.3)--(2.6,-4.3) ;
    \draw[snake=snake,->,color=red] (2.3,-4.4)--+(.2,-.4);
    \node[color=red] at (1.9,-4.6){\tiny $-Q(1)$};
    \node at (-1,-3.6) {Measure}; 
    \node at (2.2,-3.6) {Erasure}; 
\end{tikzpicture}
%
\caption{(Color online) a) As the lead enters in contact with the quantum dot to be erased, an imperfect measurement is performed. b) The time dependent protocol $\mu(t)$ applied during the contact time $t$ depends on the output of that measurement. \label{fig:dotdeviceFeedback}}
\end{figure}
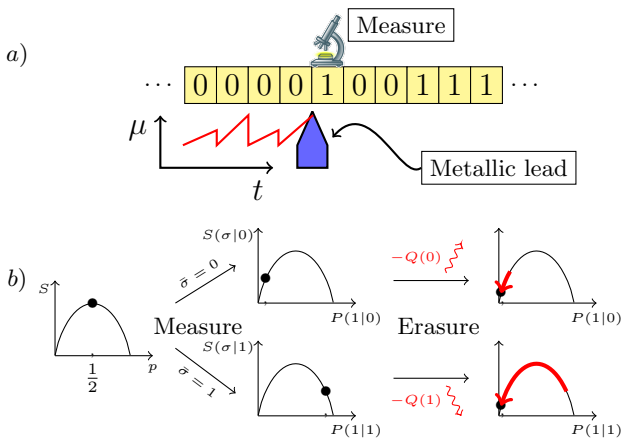

We now turn to an erasure process assisted by the feedback process depicted in Fig.~\ref{fig:dotdeviceFeedback}. An imperfect measurement is performed on the bit to be erased and the ensuing protocol depends on the output of that measurement. 

The two possible states of the bit, empty and filled, are denoted by $\sigma=0,1$ and the probability to find the bit in state $\sigma$ at time $t$ is denoted $P_t(\sigma)$. We consider {\it ideal measurements} which do not perturb the system measured. Therefore, the probability to find the bit in a given state $\sigma$ remains the same right after as right before the measurement which occurs at time $t=0$: $P_0(\sigma=1)=p_i$ and $P_0(\sigma=0)=1-p_i$. The two possible outcomes of the measurement are denoted by $\bar\sigma=0,1$. The accuracy of the measurement apparatus is characterized by the conditional probability $P_{0}(\sigma|\bar\sigma)$ to find the bit in state $\sigma$ when the measurement outcome $\bar\sigma$ is realized:
\begin{equation} \label{eq:initial}
P_{0}(\sigma|\bar\sigma)=\left\lbrace
\begin{array}{c l}
1-\delta\quad & \mathrm{\it if}\; \sigma=\bar\sigma \\
\delta \quad  & \mathrm{\it if }\; \sigma\neq \bar\sigma
\end{array}\right. .
\end{equation} 
A {\it perfect measurement} corresponds to $\delta=0$ and fully characterizes the system state while a {\it useless measurement} corresponds to $\delta=1/2$ and does not provide any additional information about the system state. After the measurement, the protocol $\epsilon(\bar\sigma)$ depends on the measurement outcomes. This means that the system will be described by the conditional probabilities $P_t(\sigma|\bar\sigma)$ which evolve according to the master equation (\ref{eq:model}) with the initial condition (\ref{eq:initial}).
The joint probability $P_t(\sigma,\bar\sigma)$ is related to the time-dependent conditional probability by
\begin{equation}
P_t(\sigma,\bar\sigma)= P_t(\sigma|\bar\sigma) P(\bar\sigma) .\label{DefCondProb}
\end{equation}
It corresponds to the Shannon entropy
\begin{eqnarray}
S_t(\sigma,\bar\sigma) &=& -\sum_{\sigma,\bar\sigma} P_t(\sigma,\bar\sigma) \ln P_t(\sigma,\bar\sigma)\nonumber\\
&=& S(\bar\sigma)+\sum_{\bar\sigma} P(\bar\sigma) S_t(\sigma|\bar\sigma) , \label{eq:relative}
\end{eqnarray}
where in the second line we defined 
\begin{eqnarray}
S(\bar\sigma) &=& -\sum_{\bar\sigma} P(\bar\sigma) \ln P(\bar\sigma) \\
S_t(\sigma|\bar\sigma) &=& -\sum_\sigma P_t(\sigma|\bar\sigma) \ln P_t(\sigma|\bar\sigma) \label{condShannon}.
\end{eqnarray}
The probability to measure an output $\bar\sigma$, $P(\bar\sigma)$ in (\ref{DefCondProb}), is obtained from the condition
\begin{eqnarray}
P_0(\sigma) = \sum_{\bar\sigma} P_0(\sigma,\bar\sigma) = \sum_{\bar\sigma} P_0(\sigma|\bar\sigma) P(\bar\sigma) .
\end{eqnarray}
We easily find that for our model
\begin{eqnarray}
P(\bar\sigma=1) = 1 - P(\bar\sigma=0) = \frac{p_i-\delta}{1-2\delta}. \label{ProbOut}
\end{eqnarray}
The Shannon entropy of the bit at time $t$, 
\begin{eqnarray}
S_t(\sigma) = - \sum_{\sigma} P_t(\sigma) \ln P_t(\sigma),
\end{eqnarray}
is related to the mutual information between the system and the measurement outcome by \cite{CoverThomas}
\begin{eqnarray} \label{eq:mutual}
M_t &=& S(\bar\sigma)+S_t(\sigma)-S_t(\sigma,\bar\sigma) \\
&=& \sum_{\sigma,\bar\sigma} P_t(\sigma,\bar\sigma) \ln \frac{P_t(\sigma,\bar\sigma)}{P_t(\sigma) P(\bar\sigma)} \geq 0. \nonumber 
\end{eqnarray}
By combining Eqs.~(\ref{eq:relative}) and (\ref{eq:mutual}), the change in the bit entropy can be written as 
\begin{eqnarray}
\Delta S_t = \Delta M_t +\sum_{\bar\sigma} P(\bar\sigma) \Delta S_t(\sigma|\bar\sigma) . \label{EntChangMut}
\end{eqnarray}
Since $P_t(\sigma|\bar\sigma)$ evolves according to Eq.~(\ref{eq:model}) with protocol $\epsilon(\bar\sigma)$, using traditional stochastic thermodynamics at the level of this conditional probability, we find that 
\begin{eqnarray}
\Delta S_t(\sigma|\bar\sigma) = Q(\bar\sigma)/T + \Delta_{\rm \bold i} S(\bar\sigma) \label{Modif2ndLaw} ,
\end{eqnarray}
where $Q(\bar\sigma)$ and $\Delta_{\rm \bold i} S(\bar\sigma) \geq 0$ are the heat and the entropy production associated to the dynamics following a measurement output $\bar\sigma$. We can thus rewrite (\ref{EntChangMut}) as 
\begin{eqnarray}
\Delta S_t = \Delta M_t + Q_F/T + \langle \Delta_{\rm \bold i} S \rangle, \label{eq:second}
\end{eqnarray}
where $\Delta M_t$ is the change in mutual information and 
\begin{eqnarray}
&& Q_F = \sum_{\bar\sigma} P(\bar\sigma) Q(\bar\sigma) \\
&& \langle \Delta_{\rm \bold i} S \rangle = \sum_{\bar\sigma} P(\bar\sigma) \Delta_{\rm \bold i} S(\bar\sigma) \geq 0 \nonumber
\end{eqnarray}
are respectively the heat and entropy production averaged over the possible measurement output giving rise to different protocols.
Eq.~(\ref{eq:second}) can be seen as a generalization of the second law of stochastic thermodynamics (\ref{eq:2law}) in absence of feedback to situations with feedback. It imposes the following bound on the heat released by the system
\begin{equation}\label{eq:land}
-Q_F \ge T \Delta M_t - T \Delta S_t \ge -Q_{F,\mathrm{min}},
\end{equation}
where the minimum heat release possible is given by 
\begin{eqnarray}
-Q_{F,\mathrm{min}} \equiv - T M_{0} - T \Delta S_t .\label{MinHeatFeed}
\end{eqnarray}
In order for the released heat $-Q_F$ to reach the intermediate bound in (\ref{eq:land}), all the erasure processes following the measurement have to be performed quasistatically: $\Delta_{\rm \bold i} S(\bar\sigma)=0$ for all $\bar\sigma$. The bound $-Q_{F,\mathrm{min}}$ can only be reached if in addition all the different protocols end up at time $t$ at a same final value independently on the measurement outputs. Indeed, since the probability of a quasistatic processes is fully determined by its protocol, at the end of the each process we would have that $P_t(\sigma|\bar\sigma)=P_t(\sigma)$, and thus that all the initial mutual information has been consumed at the end of the process: $M_t=0$. 

We will assume, as we did in last section, that $p_i=1/2$. Therefore, using (\ref{ProbOut}), we have $P(\bar\sigma=1) = 1 - P(\bar\sigma=0)$ $= 1/2$. With the measurement errors (\ref{eq:initial}) and using (\ref{ProbOut}) and (\ref{eq:mutual}), the initial mutual information becomes 
\begin{eqnarray}\label{MutalIni}
&&M_{0} = \ln 2 - S_\delta ,\\ 
&&S_\delta \equiv - \delta \ln \delta - (1-\delta) \ln(1-\delta).\nonumber
\end{eqnarray}
We have seen in last section that perfect erasure can be achieved in the quasistatic limit. This means that for each of the two measurement output, we can reach $P_t(\sigma=1|\bar\sigma)=P_t(\sigma=1)=0$ so that $\Delta S_t=-\ln 2$ and $M_t=0$. As a result, the minimum possible heat released in presence of feedback is given by
\begin{equation}
-Q_{F,\mathrm{min}} = T \ln 2- T M_{0} = T S_\delta . \label{MinHeatFeedModel}
\end{equation}
This result can be viewed as an extension of the Landauer principle in presence of a feedback. In the limit of a perfect measurement ($\delta=0$) the minimal heat released completely vanishes. 

We turn now to the protocols $\epsilon_t(\bar\sigma)$ that minimize the released heat required to erase a given amount of information in finite time. To make the comparison with last section meaningful, we impose the same change in Shannon entropy with and without feedback. We will therefore minimize the heat released in going from $P_0(\sigma=1)=p_i=1/2$ to $P_t(\sigma=1)=p_f$ in a finite time $t$. The final values of the conditional probabilities $P_t(\sigma=1|\bar\sigma)$ have to satisfy the constraint 
\begin{equation} \label{eq:systemprob}
P_t(\sigma) = \sum_{\bar\sigma} P_t(\sigma|\bar\sigma) P(\bar\sigma) .
\end{equation}
To simplify the notation we define 
\begin{equation} 
p^{(\bar\sigma)}_f \equiv P_t(\sigma=1|\bar\sigma)=1-P_t(\sigma=0|\bar\sigma),
\end{equation}
Since $P(\bar\sigma)=1/2$, we find that 
\begin{equation} \label{eq:relationpf}
p_f^{(0)}=2p_f-p_f^{(1)}.
\end{equation}
This fixes $p_f^{(0)}$ in terms of $p_f$ but leaves $p_f^{(1)}$ free. 
The average heat released now reads
\begin{equation} \label{eq:Qfeedback}
Q_F= \sum_{\bar\sigma} P(\bar\sigma) Q(\bar\sigma)=\frac{1}{2} \big( Q(0)+Q(1) \big).
\end{equation}
Minimizing $-Q_F$ for a given time $t$ is done by first separately finding the two optimal protocols $\epsilon_t(0)$ and $\epsilon_t(1)$ minimizing $Q(0)$ and $Q(1)$ respectively, using the final probabilities $p_f^{(0)}$ and $p_f^{(1)}$ which are related by (\ref{eq:relationpf}). The second step consists in further minimizing the resulting expelled heat $-Q_F$ with respect to $p_f^{(1)}$. The minimum is reached for $p_{f,\mathrm{opt}}^{(1)}$. 

In absence of feedback we have seen that to avoid divergences in the heat we need to fulfill the condition $2 p_f > \exp{\{-t\}}$. Similarly in presence of feedback, to avoid divergences in the heat $Q(0)$ and $Q(1)$, we need to satisfy $p_{f}^{(1)} > (1-\delta) \exp{\{-t\}}$ and $p_{f}^{(0)} > \delta \exp{\{-t\}}$ respectively. Using (\ref{eq:relationpf}), these two conditions combine as
\begin{equation} \label{eq:window}
(1-\delta) \exp{\{-t\}} < p_{f}^{(1)} < 2 p_f - \delta \exp{\{-t\}}.
\end{equation}
In the limit $2 p_f \rightarrow \exp{\{-t\}}$ where the heat of the process without feedback becomes divergent, the bounds collapse and all the critical final probabilities become identical. For $2 p_f > \exp{\{-t\}}$, $p_{f,\mathrm{opt}}^{(1)}$ will be located within the bounds (\ref{eq:window}) where none of the critical final probabilities are reached.

\begin{figure}[t]
\begin{tikzpicture}
\node {\includegraphics[scale=0.33]{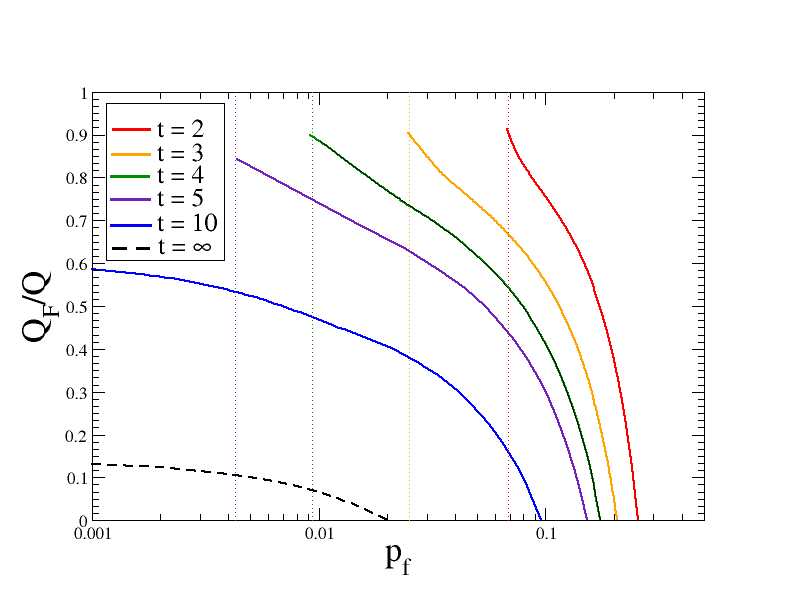}};
\end{tikzpicture}
\caption{(Color online) Ratio $Q_F/Q$ as a function of the final probability $p_f$ for different erasure times $t$. The curves bend down as $t$ increases and the dashed curve corresponds to the quasistatic limit (\ref{eq:quasistaticQFQ}). The vertical lines denote the critical probability $p_c$ reachable for each erasure time $t$. The measurement error is fixed at $\delta=0.02$. \label{fig:ratios_p}}
\end{figure} 
\begin{figure}[b]
\includegraphics[scale=0.33]{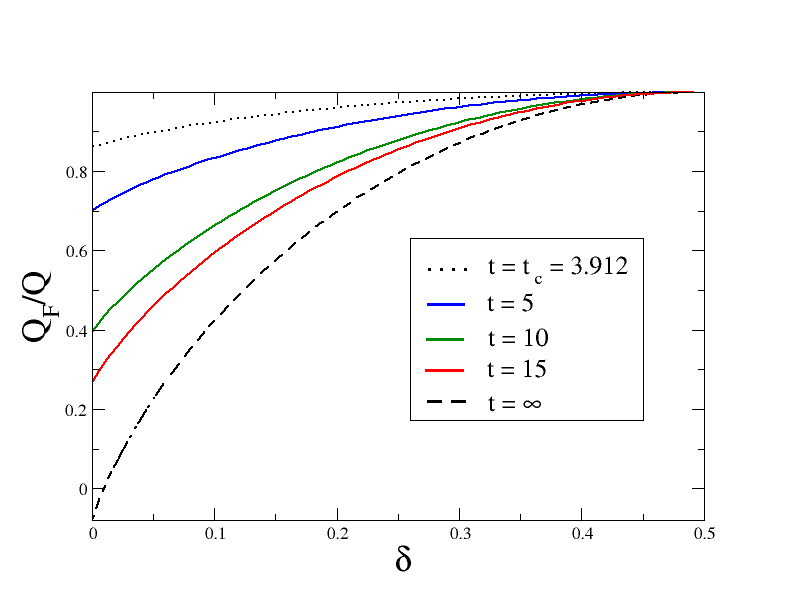}
\caption{(Color online) Ratio $Q_F/Q$ as a function of the measurement error $\delta$ for different erasure times $t$. The final probability is fixed at $p_f=0.01$. The dotted curve represents the ratio at the critical time $t_\mathrm{min}=-\ln 2 p_f$ while the dashed curve corresponds to the quasistatic limit (\ref{eq:quasistaticQFQ}). \label{fig:ratios_eps}}
\end{figure}

In the quasistatic limit, we can derive the minimal value of $-Q_F$ predicted by (\ref{MinHeatFeed}) with (\ref{MutalIni}) explicitly. Using (\ref{Modif2ndLaw}) with (\ref{condShannon}), we get that
\begin{eqnarray}
Q_F&=&\frac{1}{2}\left(-p_f^{(1)}\ln p_f^{(1)}-(1-p_f^{(1)})\ln(1-p_f^{(1)})-S_\delta\right) \nonumber \\
&&+\frac{1}{2}\left(-p_f^{(0)}\ln p_f^{(0)}-(1-p_f^{(0)})\ln(1-p_f^{(0)})-S_\delta\right). \nonumber \\
\end{eqnarray}
If we impose the constraint (\ref{eq:relationpf}), the derivative of $Q_F$ with respect to $p_f^{(1)}$ vanishes for $p_{f,\mathrm{opt}}^{(1)}=p_{f,\mathrm{opt}}^{(0)}=p_f$. Since in the quasistatic limit, $-Q_F/T = \Delta M_t-\Delta S_t$, by minimizing the heat at fixed $\Delta S$, we are in fact minimizing the final mutual information $M_t$ which vanish precisely when $p_{f}^{(0)}=p_{f}^{(1)}$. Using (\ref{MutalIni}), the minimum value of $-Q_F$ then reads
\begin{equation}
-Q_{F,\mathrm{min}}=- T M_0 - T \Delta S_t = T S_\delta - T S_f. \label{QuasistatQF}
\end{equation}

A relevant quantity to look at is the ratio $Q_F/Q$ between the heat released in presence and in absence of the feedback to erase a given amount of information in finite time. In the quasistatic limit, using (\ref{QuasistatQF}), we find that
\begin{equation}\label{eq:quasistaticQFQ}
\frac{Q_{F,\mathrm{min}}}{Q_{\mathrm{min}}}=\frac{S_\delta-S_f}{\ln2-S_f}.
\end{equation}
Since $S_{\delta} \leq \ln 2$, except for useless measurements ($\delta=1/2$), the feedback always helps to reduce the expelled heat.
In finite time, the ratio $Q_F/Q$ corresponding to the protocols minimizing heat has been calculated in Fig.~\ref{fig:ratios_p}, for a given measurement accuracy, as a function of the final probability $p_f$ and of the erasing time $t$. This ratio is always lower than one and decreases for longer erasure times. This shows that the reduction in the expelled heat thanks to the feedback increases as the contact time increases and is most significant in the quasistatic limit. The same ratio has been calculated in Fig.~\ref{fig:ratios_eps}, for a fixed value of the final probability $p_f=0.01$, as a function of the measurement error $\delta$ and of the erasing time $t$. As we approach the useless measurements limit $\delta \to 1/2$, the ratio goes to one independently of the erasure time indicating no gain by the feedback. In the perfect measurement limit $\delta \to 0$, as the contact time becomes longer the ratio approaches zero and may even become negative. This happens in the $t \to \infty$ limit when $S_\delta < S_f$ as predicted by (\ref{QuasistatQF}).  

\section{Conclusions} \label{section:conclu}

We proposed in this paper a model to study the thermodynamics of information erasure in finite time. It consists of an array of single level quantum dots which can store classical information given that each dot with its two states, with or without an electron, constitutes a classical bit. The initial information is quantified by the Shannon entropy $S_i$. The erasure process consists in decreasing that information by $-\Delta S=S_i-S_f>0$ and is performed by a metallic lead moving at constant speed along the array of quantum dots. During each lead-dot contact time $t$ a time dependent protocol controls the lead chemical potential.

In the first part of the paper we considered the situation where the same protocol is applied to each dot. We found the following results. 
Perfect erasure $S_f=0$ of any finite initial information $S_i$ always requires an infinitely long contact time $t \to \infty$. 
Imperfect erasure (which leaves some errors $S_f$) with a maximal erasing efficiency $\eta=1$ requires a quasistatic protocol and thus again a diverging contact time. The erasing efficiency $\eta$ measures the fraction of the generated heat (divided by temperature) ($-Q/T$) that is used to erase the information while the rest is lost as entropy production. 
Reaching Landauer's lower bound $S_i=-Q/T$ requires perfect erasure with efficiency one. 
In finite time, errors and entropy production are unavoidable. 
Attempting to minimize errors will result in a diverging heat and a vanishing erasing efficiency. For larger values of the error, we calculated the protocols minimizing the heat generation and studied the resulting heat generation and efficiency as a function of the error and the erasure time. We studied the behavior of the erasing power which characterizes the average erased information per unit time achieved by generating a given amount of heat. We finally studied the erasing efficiency at maximum erasing power.   

In the second part of the paper we considered information erasure with help of a feedback process. In this case the protocol applied on each dot depends on the output of an imperfect measurement of the state of the dot which occurs as soon as the lead enters in contact with the dot. We showed that the measurement creates mutual information between the system and the measurement outcomes which lowers the expelled heat obtained in absence of feedback. The lower bound continues to be reached in infinite time for quasistatic protocols and can be well below Landauer's limit. In case of perfect measurements this bound even vanishes. In finite time the lower bound cannot be reached and additional heat gets released. We studied in detail the proportion by which the heat released to erase a given amount of information in finite time is reduced by the feedback.        

The field of finite time information processing is still largely unexplored despite its conceptual and technological importance. This study shows that stochastic thermodynamics provides a useful conceptual framework to make significant progress in this direction.

\section{Acknowledgements}

G.D. and M.E. are supported by the National Research Fund, Luxembourg in the frame of project FNR/A11/02. G. B. B. thanks YOK (The Council of Higher Education in Turkey) for its financial support.

%
\end{document}